\documentclass[amstex,twocolumn,superscriptaddress,showpacs,floats,floatfix,aps,pra]{revtex4}
\usepackage{amssymb}
\usepackage{amsmath}
\usepackage{calc}
\usepackage{graphicx}
\usepackage{bm}

\def\ket#1{|#1\rangle}
\def\ket#1{#1}
\def\tsum{\sum\nolimits}
\def\rms{\textsc{rms}}
\def\d{\mathbf{d}}
\def\b{\mathbf{b}}

\begin{document}

\title{Achromatic multiple beam splitting by adiabatic passage in optical waveguides}

\begin{abstract}
A novel variable achromatic optical beam splitter with one input and $N$ output waveguide channels is introduced. 
The physical mechanism of this multiple beam splitter is adiabatic passage of light between neighboring optical waveguides
 in a fashion reminiscent of the technique of stimulated Raman adiabatic passage in quantum physics.
The input and output waveguides are coupled via a mediator waveguide and the ratios of the light intensities in the output channels
 are controlled by the couplings of the respective waveguides to the mediator waveguide.
Due to its adiabatic nature the beam splitting efficiency is robust to variations in the experimental parameters.
\end{abstract}

\author{Andon A. Rangelov}
\affiliation{Department of Physics, Sofia University, 5 James Bourchier blvd, 1164 Sofia, Bulgaria}
\affiliation{Corresponding author: rangelov@phys.uni-sofia.bg}
\author{Nikolay V. Vitanov}
\affiliation{Department of Physics, Sofia University, 5 James Bourchier blvd, 1164 Sofia, Bulgaria}
\date{\today }

 \maketitle



An optical beam splitter is a device that splits a beam of light normally into two parts with the same, or different intensities.
The most common form of a beam splitter is a rectangle glass prisms that reflects half of the light and transmits the other half due to frustrated total internal reflection. 
Traditional beam splitters suffer from monochromatic limitation (each is designed for a specific wavelength) and they are sensitive to the angle of incidence.
Recently, a waveguide structure that is able to divide an input signal into two equal outputs for a broadband spectrum was proposed as a new kind of achromatic beam splitter \cite{Dreisow2009A};
 it is analogous to the technique of fractional stimulated Raman adiabatic passage (STIRAP) \cite{fSTIRAP} in quantum optics.

The analogies with two-, three- and many-state quantum systems is a hot topic in the field of waveguide structures,
 which have led to implementations of some well-known techniques from quantum physics to control of light propagation in optical waveguides.
These analogues include Rabi oscillations \cite{Longhi 2005A},
 Landau--Zener tunneling \cite{Khomeriki,Longhi 2005},
 STIRAP \cite{Paspalakis,Longhi 2006,Longhi 2007,Lahini,Dreisow2009A},
 fractional STIRAP \cite{Paspalakis,Dreisow2009B},
 STIRAP extensions over many states \cite{Valle,Tseng},
 and STIRAP via the continuum \cite{Longhi2008,Dreisow2009C,Longhi2009A}.
Research in this direction is still growing rapidly, as reviewed recently \cite{Longhi 2009B}.

In this paper, we propose a variable achromatic optical beam splitter with one input and $N$ output waveguide channels, which are connected via a mediator waveguide. 
The device is an analogue to tripod STIRAP \cite{Theuer,Unanyan,Vewinger} in quantum optics,
 and hence it should enjoy the same advantages as STIRAP in terms of efficiency and robustness against variations of the experimental parameters,
 such as the waveguides couplings, the distance between the waveguides and their geometry.

\begin{figure}[t]
\includegraphics[width=0.95\columnwidth]{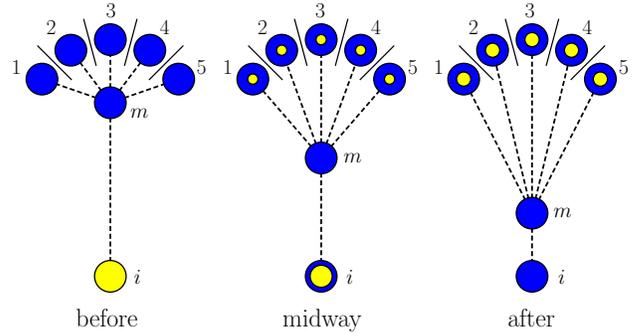}
\caption{(Color online) Cross section of a possible waveguide arrangement of the multiple beam splitter with an input channel $\ket{i}$, mediator waveguide $\ket{m}$ and $N=5$ output channels
 in the beginning (left), midway (middle) and in the end (right).
The dashed lines depict the couplings between the waveguides.
The light spots indicate the waveguides wherein the light is found and the spot area is proportional to the corresponding light intensity.
}
\label{Fig-fan}
\end{figure}

\begin{figure}[t]
\centerline{\includegraphics[width=0.95\columnwidth]{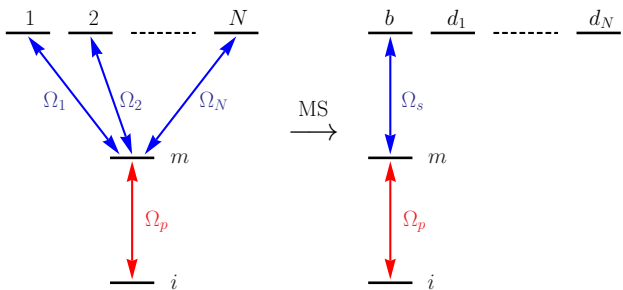}}
\caption{(Color online) 
Linkage pattern of the waveguides (left) and the effective couplings after the Morris-Shore transformation (right).}
\label{Fig2}
\end{figure}

We consider a system of one input, one intermediate and $N$ output waveguides, as shown in Fig.~\ref{Fig-fan}.
In the paraxial approximation with weakly curved waveguides the propagation of monochromatic light is described by a set of $N+2$ coupled differential equations (in matrix form) \cite{Dreisow2009A,Longhi 2005,Paspalakis,Longhi 2006,Longhi 2007},
\begin{equation}\label{three states system}
i\frac{d\mathbf{A}(z)}{dz}=\mathbf{H}(z)\mathbf{A}(z),
\end{equation}
where $\mathbf{A}=\left[ a_i(z), a_m(z), a_{1}(z),a_{2}(z),\ldots,a_{N}(z)\right] ^{T}$ and
\begin{equation}
\mathbf{H}(z) = \left[\begin{array}{ccccc}
0 & \Omega _{p}(z) & 0 & \cdots & 0 \\
\Omega _{p}(z) & 0 & \Omega _{1}(z) & \cdots & \Omega _{N}(z) \\
0 & \Omega _{1}(z) & 0 & \cdots & 0 \\
\vdots & \vdots & \vdots & \ddots & \vdots \\
0 & \Omega _{N}(z) & 0 & \cdots & 0%
\end{array}\right] .  \label{N-pod Hamiltonian}
\end{equation}
Here $a_{k}(z)$ ($k=i,m,1,2,\ldots,N$) is the light amplitude in the $k$-th waveguide and the corresponding light intensity is $I_{k}=|a_{k}(z)|^{2}$. 
The coupling coefficient between waveguides $\ket{i}$ and $\ket{m}$ is $\Omega _{p}(z)$ and the one between waveguides $\ket{m}$ and $\ket{k}$ is $\Omega_{k}(z)$ ($k=1,2,\ldots,N$).
We assume that the latter couplings share the same time dependence but they may have in general different magnitudes.
The subscript $p$ in the coupling $\Omega_p(z)$ is introduced in anticipation of the forthcoming analogy to the pump field in STIRAP.
The waveguides of the output channels should not be coupled to each other but only to the mediator waveguide $\ket{m}$;
 hence they are supposed to be isolated from each other, as sketched in Fig.~\ref{Fig-fan}.
The null diagonal elements in the matrix $\mathbf{H}(z)$ are due to the assumption of nearest-neighbor tight-binding approximation;
 hence only the direct coupling between the waveguides is considered (see Fig.~\ref{Fig2}).

If we map the coordinate dependence into a time dependence, Eq.~\eqref{N-pod Hamiltonian} is identical to the time-dependent Schr\"{o}dinger equation;
 the vector $\mathbf{A}(z)$ and the driving matrix $\mathbf{H}(z)$ correspond to the quantum state vector and the Hamiltonian, respectively.
In the absence of losses the quantity $|\mathbf{A}(z)|^{2} = |a_i(z)|^2 + |a_m(z)|^2 + \sum_{k=1}^{N}|a_{k}(z)|^{2}$ is conserved, like the total population in a coherently driven quantum system.

The physical mechanism of the proposed beam splitter is most easily explained by the so-called Morris-Shore (MS) transformation \cite{MS}, which is illustrated in Fig.~\ref{Fig2}.
This transformation is applied on the upper set of transitions between the middle waveguide $\ket{m}$ and the output waveguides $\ket{k}$ ($k=1,2,\ldots,N$), which form an N-pod (or ``fan'') linkage pattern \cite{Kyoseva}.
The MS basis comprises the mediator waveguide $\ket{m}$, a set of decoupled (dark) superpositions of output waveguides $\ket{\d_k}$ ($k=1,2,\ldots,N-1$) and a bright superposition $\b$ of output waveguides, which is coupled to the mediator $\ket{m}$ with a coupling which is the root-mean-square (\rms) of the initial couplings,
\begin{equation}
\Omega_s(z) = \sqrt{\tsum_{k=1}^{N}\Omega_k(z)^2}.
\end{equation}
The decoupled superpositions $\ket{\d_k}$ are of no significance in the present context.
The bright superposition has the vector form
\begin{equation}\label{bright}
\b(z) = \frac{[0,0,\Omega_1(z),\Omega_2(z),\ldots,\Omega_N(z)]^T}{\Omega_s(z)},
\end{equation}
i.e. it does not contain contributions from the input waveguide $\ket{i}$ and the middle waveguide $\ket{m}$.
The output-waveguide components are proportional to the respective couplings $\Omega_k(z)$;
 if these couplings are equal then the bright superposition $\b(z)$ will be an equally-weighted superposition of all output waveguides.

The three-waveguide ladder $\ket{i}\to\ket{m}\to\b$ in the MS basis is the subspace in which the beam splitting takes place via STIRAP-like adiabatic transfer of light from $\ket{i}$ to $\b$.
To this end, the waveguides must be arranged in such a manner that the \rms~``Stokes'' coupling $\Omega_s(z)$ precedes the ``pump'' coupling $\Omega_p(z)$ but the two must overlap partly, as in STIRAP.
A cross section of the possible geometry of the waveguides is shown in Fig.~\ref{Fig-fan}.
Because of the evanescent-wave nature of the waveguide couplings, which depend strongly on the distance between the waveguides, the ``Stokes'' couplings dominate over the ``pump'' coupling in the beginning and vice versa in the end.

As in STIRAP, the adiabatic passage of light proceeds via the dark superposition \cite{Thanopoulos},
\begin{equation}
\label{dark state}
\d(z) = \frac{[\Omega_s(z),0,-\Omega_p(z)]^T}{\sqrt{\Omega_p(z)^2+\Omega_s(z)^2}} ,
\end{equation}

\begin{figure}[tb]
\centerline{\includegraphics[width=0.8\columnwidth]{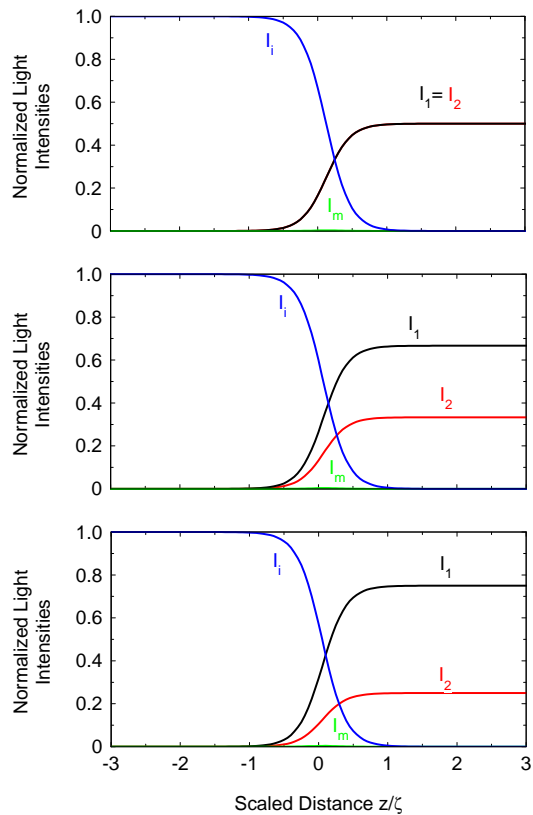}}
\caption{(Color online) Variable two-beam splitting between waveguides $\ket{1}$ and $\ket{2}$.
\emph{Top frame:}    1:1 beam splitting for $\Omega_{1}(z) =\Omega_{2}(z) = \Omega_0 f_s(z)$.
\emph{Middle frame:} 2:1 beam splitting for $\Omega_{1}(z) = \Omega_0 f_s(z)$ and $\Omega _{2}(z) = (\Omega_0/\sqrt{2}) f_s(z)$.
\emph{Bottom frame:} 3:1 beam splitting for $\Omega_{1}(z) = \Omega_0 f_s(z)$ and $\Omega _{2}(z) = (\Omega_0/\sqrt{3}) f_s(z)$.
In all frames,  $\Omega_p(z) = \Omega_0 f_p(z)$, $f_p(z)=\exp[-(z/\zeta-0.7)^2]$, $f_s(z)=\exp[-(z/\zeta+0.7)^2]$, and $\Omega_0=25/\zeta$.
Here $\zeta$ is the characteristic length of the setup; it is used as the unit for length and $1/\zeta$ as the unit for coupling.
}
\label{Fig3}
\end{figure}

By definition, in the adiabatic limit the system stays in an eigenvector of the Hamiltonian \cite{VitanovA,VitanovB}.
If $\Omega_s(z)$ occurs before $\Omega_p(z)$, as we assume, then the dark superposition $\d(z)$ is equal to the input waveguide initially;
 hence if adiabatic evolution is enforced the light will remain in the dark superposition $\d(z)$ all along.
In the end, the dark superposition $\d(z)$ is equal to the bright output superposition $\b$ and hence, the input beam is split into $N$ components; multiple beam splitting has occurred.
The ratios of the intensities in the output waveguides are determined by the squared values of the respective couplings to the mediator waveguide, as it follows from Eq.~\eqref{bright}.
It is important that the final intensities in the output waveguide channels do not depend on the precise form of the couplings, but only on the final ratios of the individual couplings. 
Moreover, the mediator waveguide $\ket{m}$ receives no light intensity throughout the beam splitting because it has no component in the dark superposition \eqref{dark state}.

Figure \ref{Fig3} shows numerical simulations of the light intensities in the input and output waveguides for $N=2$ output waveguides.
The middle waveguide has only negligibly small light intensity because the evolution is almost adiabatic.
In all cases the light is transferred adiabatically from the input waveguide $\ket{i}$ to the output waveguides $\ket{1}$ and $\ket{2}$ with nearly 100\% efficiency.
The splitting ratio is controlled by the squares of the respective couplings: it is 1:1 in the upper frame, 2:1 in the middle frame, and 3:1 in the bottom frame.

\begin{figure}[tb]
\centerline{\includegraphics[width=0.8\columnwidth]{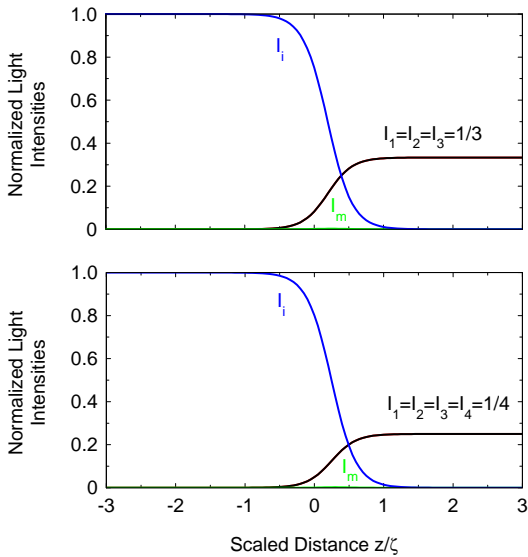}}
\caption{(Color online) Multiple beam splitting.
\emph{Top frame:} equal beam splitting between $N=3$ waveguides $\ket{1}$, $\ket{2}$, and $\ket{3}$.
\emph{Bottom frame:} equal beam splitting between $N=4$ waveguides $\ket{1}$, $\ket{2}$, $\ket{3}$, and $\ket{4}$.
In all frames,  $\Omega_p(z) = \Omega_0 f_p(z)$, $\Omega_k(z) = \Omega_0 f_s(z)$ ($k=1,2,\ldots,N$), $f_p(z)=\exp[-(z/\zeta-0.7)^2]$, $f_s(z)=\exp[-(z/\zeta+0.7)^2]$, and $\Omega_0=25/\zeta$.
}
\label{Fig4}
\end{figure}

Figure \ref{Fig4} demonstrates multiple beam splitters, in which the incident light is transferred from the input waveguide to several output waveguides.
Again, if needed, variable beam splitting can be achieved by the ratios of the couplings from the mediator to the output waveguides.

Because the physical mechanism of the beam splitter described above is based upon adiabatic evolution, which is insensitive to variations in the couplings,
 this beam splitter is expected to be achromatic, as in the beam splitter using fractional STIRAP \cite{Dreisow2009B}.

In conclusion, we have proposed a novel type of achromatic optical beam splitter with one input and $N$ output waveguide channels.
This beam splitter uses STIRAP-like adiabatic passage of light and it is therefore expected to be robust against variations of the couplings and the waveguide geometry.

This work is supported by the European Commission project FASTQUAST, and the Bulgarian National Science Fund grants D002-90/08 and DMU-03/103.


\end{document}